\documentclass[11pt]{article}
\usepackage{fullpage}
\usepackage{amsmath,amssymb}    % AMS
\usepackage{url}
\usepackage{color}
\definecolor{ForestGreen}{rgb}{0.1333,0.5451,0.1333}
\usepackage[colorlinks,linkcolor=ForestGreen,citecolor=ForestGreen,backref]{hyperref}
\usepackage{mathrsfs}           % \mathscr
\usepackage{enumerate}

\usepackage{verbatim}           % comment environment
\usepackage{xspace}             % optional space
\usepackage{graphicx,float}     % figures

\newtheorem{theorem}{Theorem}[section]

\newtheorem{corollary}[theorem]{Corollary}

\newtheorem{lemma}[theorem]{Lemma}

\newcommand{\Proof}[0]{\smallskip\noindent\textit{\textbf{Proof:}}\quad}

\newcommand{\Proofof}[1]{\smallskip\noindent\textit{\textbf{Proof of #1:}}\quad}

\newcommand{\Oh}{\ensuremath{\mathcal{O}}}

\newcommand{\bv}[1]{\mathbf{#1}}

\newcommand{\QED}[0]{\hfill\ensuremath{\blacksquare}\medspace\\}

\floatstyle{ruled}
\newfloat{algo}{tbp}{lop}%[section]
\floatname{algo}{Algorithm}

\newcommand{\norm}[1]{\|#1\|}

\usepackage{float}
\floatstyle{plain}\newfloat{myfig}{t}{figs}[section]
\floatname{myfig}{\textsc{Figure}}
\floatstyle{plain}\newfloat{myalg}{H}{algs}[section]
\floatname{myalg}{}
\newcommand{\poly}[1]{\text{poly}(#1)}

\newcommand{\expct}[3]{\ensuremath{\mathop{\text{\normalfont \textbf{E}}_{#1}}_{#2}}\left[#3\right]}
\newcommand{\prob}[1]{\ensuremath{\mathop{\text{\normalfont \textbf{P}}}}\left[#1\right]}

  \usepackage{nth}
  \usepackage{intcalc}

\begin{document}

\begin{titlepage}
\def\thepage{}
\thispagestyle{empty}

\title{Online Row Sampling}

\author{\and
Michael B. Cohen\\MIT\footnote{Work supported by NSF grant CCF-1111109.}\\ \texttt{micohen@mit.edu}
\and
Cameron Musco\\MIT\footnote{Work partially
supported by NSF Graduate Research Fellowship No. 1122374, AFOSR
grant FA9550-13-1-0042 and the NSF Center for Science of Information.}\\ \texttt{cnmusco@mit.edu}
\and
Jakub Pachocki\\Carnegie Mellon University\footnote{Work supported by NSF grant CCF-1065106.}\\ \texttt{pachocki@cs.cmu.edu}
}

\maketitle

\abstract{
Finding a small spectral approximation for a tall $n \times d$ matrix $\bv A$ is a fundamental numerical primitive.
For a number of reasons, one often seeks an approximation whose rows are sampled from those of $\bv{A}$. Row sampling improves interpretability, saves space when $\bv A$ is sparse, and preserves row structure, which is especially important, for example, when $\bv A$ represents a graph.

However, correctly sampling rows from $\bv{A}$ can be costly when the matrix is large and cannot be stored and processed in memory. Hence, a number of recent publications focus on row sampling in the streaming setting, using little more space than what is required to store the outputted approximation \cite{kelner2013spectral,kapralov2014single}.

Inspired by a growing body of work on online algorithms for machine learning and data analysis, we extend this work to a more restrictive \emph{online} setting: we read rows of $\bv A$ one by one and immediately decide whether each row should be kept in the spectral approximation or discarded, without ever retracting these decisions. We present an extremely simple algorithm that approximates $\bv A$ up to multiplicative error $\epsilon$ and additive error $\delta$ using
$\Oh(d \log d \log (\epsilon\norm{\bv A}_2^2/\delta) / \epsilon^2)$ online samples, with memory overhead proportional to the cost of storing the spectral approximation.
We also present an algorithm that uses $\Oh(d^2)$ memory but only requires $\Oh(d \log (\epsilon\norm{\bv A}_2^2/\delta) / \epsilon^2)$ samples, which we show is optimal.

Our methods are clean and intuitive, allow for lower memory usage than prior work, and expose new theoretical properties of leverage score based matrix approximation.
}

\end{titlepage}

\section{Introduction}
\label{sec:intro}

\subsection{Background}

A spectral approximation to a tall $n \times d$ matrix $\bv A$ is a smaller, typically $\tilde \Oh(d) \times d$ matrix $\bv{\tilde A}$ such that $\norm{\bv{\tilde A} \bv{x}}_2 \approx \norm{\bv{ A} \bv{x}}_2$ for all $\bv{x}$. Typically one asks for a multiplicative approximation, which guarantees that $(1-\epsilon)\norm{\bv{ A} \bv{x}}^2_2 \le \norm{\bv{\tilde A} \bv{x}}^2_2 \le (1+\epsilon)\norm{\bv{ A} \bv{x}}^2_2$. In other notation, $(1-\epsilon)\bv{A} \preceq \bv{\tilde A} \preceq (1+\epsilon) \bv{A}$.

Such approximations have many applications, most notably for solving least squares regression over $\bv A$ \cite{Clarkson:2013,cohen2015uniform}. If $\bv{A}$ is the vertex edge incidence matrix of a graph, $\bv{\tilde A}$ is a \emph{spectral sparsifier} \cite{spielman2004nearly}. It can be used to approximate effective resistances, spectral clustering, mixing time and random walk properties, and many other computations.

A number of recent papers focus on fast algorithms for spectral approximation. Using sparse random subspace embeddings \cite{Clarkson:2013,DBLP:conf/focs/NelsonN13,meng2013low}, it is possible to find $\bv{\tilde A}$ in input sparsity time, essentially by randomly recombining the rows of $\bv{A}$ into a smaller number of rows. In some cases these embeddings are not enough, as it is desirable for the rows of $\bv{\tilde A}$ to be a subset of rows sampled from $\bv{A}$. If $\bv{A}$ is sparse, this ensures that $\bv{\tilde A}$ is also sparse. If $\bv{A}$ represents a graph, it ensures that $\bv{\tilde A}$ is also a graph, specifically a weighted subgraph of the original.

It is well known that sampling $\Oh(d\log d/\epsilon^2)$ rows of $\bv{A}$ with probabilities proportional to their \emph{leverage scores} yields a $(1+\epsilon)$ multiplicative factor spectral approximation to $\bv{A}$. Further, this sampling can be done in input sparsity time, either using subspace embeddings   to approximate leverage scores, or using iterative sampling techniques \cite{li2013iterative}, some that only work with subsampled versions of the original matrix \cite{cohen2015uniform}.

\subsection{Streaming and Online Row Sampling}

When $\bv{A}$ is very large, input sparsity runtimes are not enough -- memory restrictions also become important. Hence, recent work has tackled row sampling in a streaming model of computation. \cite{kelner2013spectral} presents a simple algorithm for sampling rows from an insertion only stream, using space approximately proportional to the size of the final approximation. \cite{kapralov2014single} gives a sparse-recovery based algorithm that works in dynamic streams with row insertions and deletions, also using nearly optimal space. Unfortunately, to handle dynamic streams, the algorithm in \cite{kapralov2014single} is complex, requires additional restrictions on the input matrix, and uses significantly suboptimal runtime to recover a spectral approximation from its low memory representation of the input stream.

While the algorithm in \cite{kelner2013spectral} is simple and efficient, we believe that its proof is incomplete, and do not see an obvious way to fix it.  The main idea behind the algorithm is to sample rows by their leverage scores with respect to the stream seen so far. These leverage scores may be coarse overestimates of the true scores. However as more rows are streamed in, better estimates can be obtained and the sampled rows pruned to a smaller set.  Unfortunately, the probability of sampling a row becomes dependent on which other rows are sampled.  This seems to break the argument in that paper, which essentially claims that their process has the same distribution as would a single round of leverage score sampling.

In this paper we initiate the study of row sampling in an \emph{online setting}. As in an insertion stream, we read rows of $\bv A$ one by one. However, upon seeing a row, we immediately decide whether it should be kept in the spectral approximation or discarded, without ever retracting these decisions. 
We present a similar algorithm to \cite{kelner2013spectral}, however, since we never prune previously sampled rows, the probability of sampling a row only depends on whether previous rows in the stream were sampled. This limited dependency structure allows us to rigorously argue that a spectral approximation is obtained.

In addition to addressing gaps in the literature on streaming spectral approximation, our restricted model extends work on online algorithms for a variety of other machine learning and data analysis problems, including principal component analysis \cite{boutsidis2015online}, clustering \cite{liberty2014algorithm}, classification \cite{bordes2005huller,crammer2006online}, and regression \cite{crammer2006online}. In practice, online algorithms are beneficial since they can be highly computationally and memory efficient. Further, they can be applied in scenarios in which data is produced in a continuous stream and intermediate results must be output as the stream is processed.
Spectral approximation is a widely applicable primitive for approximate learning and computation, so studying its implementation in an online setting is a natural direction.

\subsection{Our Results}

Our primary contribution is a very simple algorithm for leverage score sampling in an online manner. The main difficultly with row sampling using leverage scores is that leverage scores themselves are not easy to compute. They are given by $l_i = \bv{a}_i^T (\bv{A}^T\bv{A})^{-1} \bv{a}_i$, and so require solving systems in  $\bv{A}^T\bv{A}$ if computed naively. This is not only expensive, but also impossible in an online setting, where we do not have access to all of $\bv{A}$.

A critical observation is that
it always suffices to sample rows by overestimates of their true leverage scores. The number of rows that must be sampled is proportional to the sum of these overestimates.
Since the leverage score of a row can only go up when we remove rows from the matrix, a
simple way to obtain an overestimate is to compute leverage score using just a subset of the other rows of $\bv{A}$. That is, letting $\bv{A}_j$ contain just $j$ of $\bv{A}$'s $n$ rows, we can overestimate $l_i$ by $\tilde l_i = \bv{a}_i^T (\bv{A}_j^T\bv{A}_j)^{-1} \bv{a}_i$
 
\cite{cohen2015uniform} shows that
if $\bv{A}_j$ is a subset of rows sampled uniformly at random, then the expected leverage score of $\bv{a}_i$ is $d/j$. This simple fact immediately gives a result for online sampling from a \emph{randomly ordered stream}.
If we compute the leverage score of the current row $\bv{a}_i$ against all previously seen rows (or some approximation to these rows), then the expected sum of our overestimates will be bounded by $d + d/2 + ... + ... + d/n = \Oh(d \log n)$. So, sampling $\Oh(d\log d \log n /\epsilon^2)$ rows will be enough obtain a $(1+\epsilon)$ multiplicative factor spectral approximation.

What if we cannot guarantee a randomly ordered input stream? Is there any hope of being able to compute good leverage score estimates in an online manner? Surprisingly the answer to this is yes - we can in fact run nearly the exact same algorithm and be guaranteed that the sum of estimated leverage scores is low, \emph{regardless of stream order}. Roughly, each time we receive a row which has high leverage score with respect to the previous rows, it must compose a significant part of $\bv{A}$'s spectrum. If $\bv{A}$ does not continue to grow unboundedly, there simply cannot be too many of these significant rows.

Specifically, we show that if we sample by the \emph{ridge leverage scores} \cite{alaoui2014fast} over all previously seen rows, which are the leverage scores computed over $\bv{A}_i^T \bv{A}_i + \lambda \bv I$ for some small regularizing factor $\lambda$, then with just $\Oh(d \log d \log (\epsilon\norm{\bv A}_2^2/\delta) / \epsilon^2)$ samples we obtain a $(1+\epsilon)$ multiplicative, $\delta$ additive error spectral approximation. That is, with high probability we sample a matrix $\bv{\tilde A}$ with $(1-\epsilon)\bv{A}^T\bv{A} - \delta\bv{I} \preceq \bv{\tilde{A}}^T\bv{\tilde{A}} \preceq (1+\epsilon)\bv{A}^T\bv{A} + \delta\bv{I}$.

To gain intuition behind this bound, note that we can convert it into a multiplicative one by setting $\delta = \epsilon \sigma_{min}(\bv{A})^2$ (as long as we have some estimate of $\sigma_{min}(\bv{A})$). This setting of $\delta$ will require taking $\Oh(d \log d \log (\kappa(\bv A)) / \epsilon^2)$ samples. If we have a polynomial bound on the condition number of $\bv{A}$, as we do, for instance, for graphs with polynomially bounded edges weights, this becomes $\Oh(d \log^2 d / \epsilon^2)$ -- nearly matching the  $\Oh(d \log d / \epsilon^2)$ achievable if sampling by true leverage scores.

Our online sampling algorithm is extremely simple. When each row comes in, we compute the online ridge leverage score, or an estimate of it, and then irrevocably either add the row to our approximation or remove it. As mentioned, it is similar in form to the streaming algorithm of \cite{kelner2013spectral}, except that it does not require pruning previously sampled rows.
This allows us to avoid difficult dependency issues. Additionally, without pruning, we do not even need to store all previously sampled rows. As long as we store a constant factor spectral approximation our previous samples, we can compute good approximations to the online ridge leverage scores. In this way, we can store just $\Oh(d \log d \log (\epsilon \norm{\bv A}_2^2/\delta) )$ rows in working memory ($\Oh(d\log^2 d)$ if we want a spectral graph sparsifier), filtering our input stream into an $\Oh(d \log d \log (\kappa(\bv A)) / \epsilon^2)$ sized output stream. Note that this memory bound in fact \emph{improves} as $\epsilon$ decreases, and regardless, can be significantly smaller than the output size of the algorithm.

In additional to our main sampling result, we use our bounds on online ridge leverage score approximations to show that an algorithm in the style of \cite{batson2012twice} allows us to remove a $\log d$ factor and sample just $\Oh(d \log (\epsilon\norm{\bv A}_2^2/\delta) / \epsilon^2)$ (Theorem \ref{thm:bss}).
This algorithm is more complex and can require $\Oh(d^2)$ working memory. However, in Theorem \ref{thm:lower_bound} we show that it is asymptotically optimal. The $\log (\epsilon\norm{\bv A}_2^2/\delta)$ factor is not an artifact of our analysis, but is truly the cost of restricting ourselves to online sampling. No algorithm can obtain a multiplicative $(1+\epsilon)$ additive $\delta$ spectral approximation taking fewer than $\Omega(d \log (\epsilon\norm{\bv A}_2^2/\delta) / \epsilon^2)$ rows in an online manner.

\section{Overview}
\label{sec:overview}
Let $\bv{A}$ be an $n \times d$ matrix with rows $\bv{a}_1, \ldots, \bv{a}_n$.
A natural approach to row sampling from $\bv A$ is picking an \emph{a priori} probability with which each row is kept,
and then deciding whether to keep each row independently.
A common choice is for the sampling probabilities to be proportional to the \emph{leverage scores} of the rows.
The leverage score of the $i$-th row of $\bv{A}$ is defined to be
\begin{align*}
    \bv{a}_i^T (\bv{A}^T\bv{A})^{\dagger} \bv{a}_i,
\end{align*}
where the dagger symbol denotes the pseudoinverse.
In this work, we will be interested in approximating $\bv{A}^T\bv{A}$ with some (very) small multiple of the identity added.
Hence, we will be interested in the \emph{$\lambda$-ridge leverage scores} \cite{alaoui2014fast}:
\begin{align*}
    \bv{a}_i^T (\bv{A}^T\bv{A} + \lambda \bv{I})^{-1} \bv{a}_i,
\end{align*}
for a parameter $\lambda > 0$.

In many applications, obtaining the (nearly) exact values of $\bv{a}_i^T (\bv{A}^T\bv{A} + \lambda \bv{I})^{-1} \bv{a}_i$ for sampling is difficult or outright impossible.
A key idea is that as long as we have a sequence $l_1, \ldots, l_n$ of \emph{overestimates} of the $\lambda$-ridge leverage scores, that is for $i=1,\ldots,n$
\begin{align*}
    l_i \geq \bv{a}_i^T (\bv{A}^T\bv{A} + \lambda \bv{I})^{-1} \bv{a}_i,
\end{align*}
we can sample by these overestimates and obtain rigorous guarantees on the quality of the obtained spectral approximation.
This notion is formalized in Theorem \ref{thm:leverage_sampling}.

\begin{theorem}
    \label{thm:leverage_sampling}
    Let $\bv{A}$ be an $n \times d$ matrix with rows $\bv{a}_1, \ldots, \bv{a}_n$.
    Let $\epsilon \in (0, 1), \delta > 0, \lambda := \delta / \epsilon, c := 8 \log d / \epsilon^2$.
    Assume we are given $l_1, \ldots, l_n$ such that for all $i = 1, \ldots, n$,
    \begin{align*}
        l_i &\geq \bv{a}_i^T (\bv{A}^T \bv{A} + \lambda \bv{I})^{-1} \bv{a}_i.
    \end{align*}
    For $i = 1, \ldots, n$, let
    $
        p_i := \min(cl_i, 1).
    $
    Construct $\bv{\tilde{A}}$ by independently sampling each row $\bv{a}_i$ of $\bv{A}$ with probability $p_i$, and rescaling it by $1 / \sqrt{p_i}$ if it is included in the sample.
    Then, with high probability,
    \begin{align*}
        (1-\epsilon)\bv{A}^T\bv{A} - \delta\bv{I} \preceq \bv{\tilde{A}}^T\bv{\tilde{A}} \preceq (1+\epsilon)\bv{A}^T\bv{A} + \delta\bv{I},
    \end{align*}
    and the number of rows in $\bv{\tilde{A}}$ is $\Oh\left(\left(\sum_{i=1}^n l_i\right) \log d / \epsilon^2\right)$.
\end{theorem}
\Proof
This sort of guarantee for leverage score sampling is well known. See for example Lemma 4 of \cite{cohen2015uniform}. If we sampled both the rows of $\bv{A}$ and the rows of $\sqrt{\lambda} \bv{I}$ with the leverage scores over $(\bv{A}^T\bv{A} + \lambda \bv I)$, we would have $(1-\epsilon)(\bv{A}^T\bv{A} + \lambda \bv I) \preceq \bv{\tilde A}^T \bv{\tilde A} \preceq(1+\epsilon)(\bv{A}^T\bv{A} + \lambda \bv I)$. However, we do not sample the rows of the identity. Since we could have sampled them each with probability $1$, we can simply subtract $\lambda \bv I = (\delta/\epsilon) \bv I$ from the multiplicative bound and have: $(1-\epsilon)\bv{A}^T\bv{A} - \delta\bv{I} \preceq \bv{\tilde{A}}^T\bv{\tilde{A}} \preceq (1+\epsilon)\bv{A}^T\bv{A} + \delta\bv{I}$.
\QED

The idea of using overestimates of leverage scores to perform row sampling has been applied successfully to various problems (see e.g. \cite{koutis2010approaching,cohen2015uniform}).
However, in these applications, access to the entire matrix is required beforehand.
In the streaming and online settings, we have to rely on partial data to approximate the true leverage scores.
The most natural idea is to just use the portion of the matrix seen thus far as an approximation to $\bv{A}$.
This leads us to introduce the \emph{online $\lambda$-ridge leverage scores}:
\begin{align*}
    l_i := \min(\bv{a}_i^T (\bv{A}_{i-1}^T \bv{A}_{i-1} + \lambda \bv{I})^{-1} \bv{a}_i, 1),
\end{align*}
where $\bv{A}_i$ $(i = 0,\ldots,n)$ is defined as the matrix consisting of the first $i$ rows of $\bv{A}$\footnote{We use the proposed scores $l_i$ for simplicity, however note that the following, perhaps more natural, definition of online leverage scores would also be effective:
\begin{align*}
    l'_i := \bv{a}_i^T (\bv{A}_{i}^T \bv{A}_{i} + \lambda \bv{I})^{-1} \bv{a}_i.
\end{align*}}.

Since clearly $\bv{A}_i^T\bv{A}_i \preceq \bv{A}^T\bv{A}$ for all $i$, it is not hard to see that $l_i$ does overestimate the true $\lambda$-ridge leverage score for row $\bv{a}_i$.
A more complex question, however, is establishing an upper bound on $\sum_{i=1}^n l_i$ so that we can bound the number of samples needed by Theorem \ref{thm:leverage_sampling}.

A core result of this work, stated in Theorem \ref{thm:online_scores}, is establishing such an upper bound; in fact, this bound is shown to be tight up to constants (Theorem \ref{thm:lower_bound}) and is nearly-linear in most cases.

\begin{theorem}
    \label{thm:online_scores}
    Let $\bv{A}$ be an $n \times d$ matrix with rows $\bv{a}_1, \ldots, \bv{a}_n$.
    Let $\bv{A}_i$ for $i \in \{0, \ldots, n\}$ be the matrix consisting of the first $i$ rows of $\bv{A}$.
For $\lambda > 0$,
let
    \begin{align*}
        l_i := \min(\bv{a}_i^T (\bv{A}_{i-1}^T \bv{A}_{i-1} + \lambda \bv{I})^{-1} \bv{a}_i, 1).
    \end{align*}
    be the online $\lambda$-ridge leverage score of the $i^{th}$ row of $\bv{A}$.
    Then
    \begin{align*}
        \sum_{i=1}^n l_i = \Oh(d \log (\norm{\bv{A}}_2^2/\lambda)).
    \end{align*}
\end{theorem}

Theorems \ref{thm:online_scores} and \ref{thm:leverage_sampling} suggest a simple algorithm for online row sampling: simply use the online $\lambda$-ridge leverage scores, for $\lambda := \delta / \epsilon$.
This produces a spectral approximation with only $\Oh(d \log d \log(\epsilon\norm{\bv{A}}_2^2/\delta)/\epsilon^2)$ rows.
Unfortunately, computing $l_i$ exactly requires us to store \emph{all} the rows we have seen in memory (or alternatively to store the sum of their outer products, $\bv{A}_i^T\bv{A}_i$).
In many cases, such a requirement would defeat the purpose of streaming row sampling.

\begin{figure}[ht]
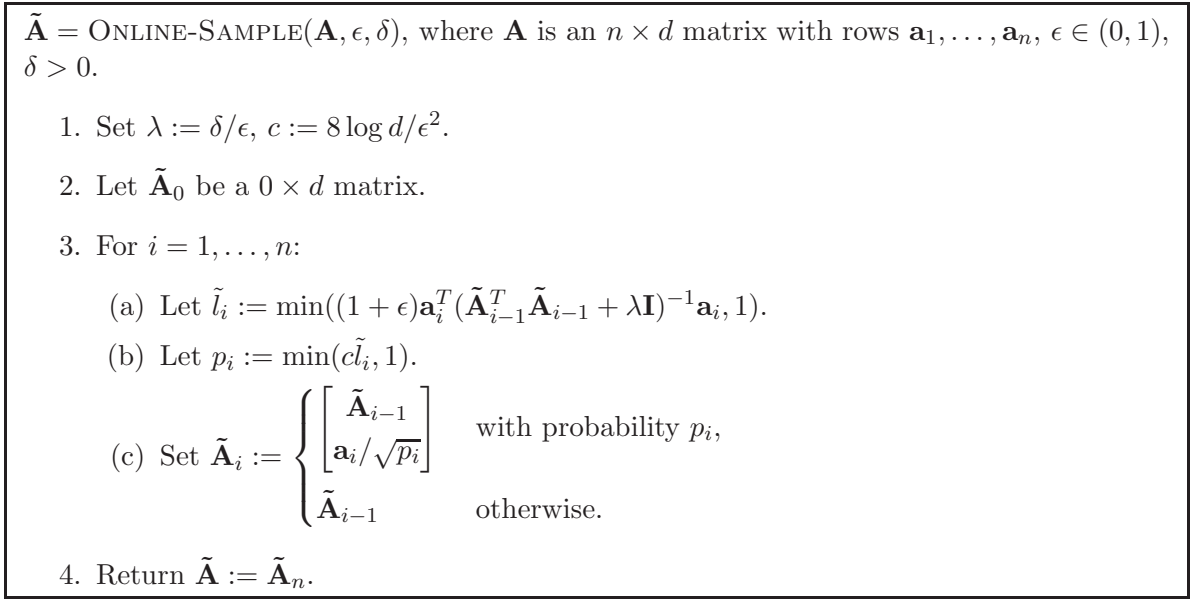

\noindent
\centering
\fbox{
\begin{minipage}{6in}
    \noindent $\bv{\tilde{A}} = \textsc{Online-Sample} (\bv{A}, \epsilon, \delta)$,
    where $\bv{A}$ is an $n \times d$ matrix with rows $\bv{a}_1, \ldots, \bv{a}_n$, $\epsilon \in (0, 1)$, $\delta > 0$.
\begin{enumerate}
\item Set $\lambda := \delta / \epsilon$, $c := 8\log d / \epsilon^2$.
\item Let $\bv{\tilde{A}}_0$ be a $0 \times d$ matrix.
\item For $i = 1, \ldots, n$:
    \begin{enumerate}
        \item Let $\tilde{l}_i := \min((1+\epsilon)\bv{a}_i^T (\bv{\tilde{A}}_{i-1}^T \bv{\tilde{A}}_{i-1} + \lambda \bv{I})^{-1} \bv{a}_i, 1)$.
        \item Let $p_i := \min(c\tilde{l}_i, 1)$.
        \item Set
            $
            \bv{\tilde{A}}_i := 
            \begin{cases}
            \begin{bmatrix} \bv{\tilde{A}}_{i-1}\\\bv{a}_i / \sqrt{p_i}\end{bmatrix} &\mbox{ with probability $p_i$,} \vspace{0.2cm}\\
                \bv{\tilde{A}}_{i-1} &\mbox{ otherwise.}\\
            \end{cases}
            $
    \end{enumerate}
\item Return $\bv{\tilde{A}} := \bv{\tilde{A}}_n$.
\end{enumerate}
\end{minipage}
}
\caption{The basic online sampling algorithm}
\label{fig:online-sample}
\end{figure}

A natural idea is to use the sample we have kept thus far as an approximation to $\bv{A}_i$ when computing $l_i$.
It turns out that the approximate online ridge leverage scores $\tilde{l}_i$ computed in this way will not always be good approximations to $l_i$; however, we can still prove that they satisfy the requisite bounds and yield the same row sample size!
We formalize these results in the algorithm $\textsc{Online-Sample}$ (Figure \ref{fig:online-sample}) and Theorem \ref{thm:online-sample}.

\begin{theorem}
    \label{thm:online-sample}
    Let $\bv{\tilde{A}}$ be the matrix returned by $\textsc{Online-Sample}(\bv{A}, \epsilon, \delta)$.
    With high probability,
    \begin{align*}
        (1-\epsilon)\bv{A}^T\bv{A} - \delta\bv{I} \preceq \bv{\tilde{A}}^T\bv{\tilde{A}} \preceq (1+\epsilon)\bv{A}^T\bv{A} + \delta\bv{I},
    \end{align*}
    and the number of rows in $\bv{\tilde{A}}$ is $\Oh(d \log d \log(\epsilon\norm{\bv{A}}_2^2/\delta)/\epsilon^2)$.
\end{theorem}

To save computation, we note that, with a small modification to our analysis, we can run $\textsc{Online-Sample}$ with batch processing of rows. Specifically, say we start from the $i^{th}$ position in the stream. we can store the next $b=\Oh(d)$ rows. We can then compute sampling probabilities for these rows all at once using a system solver for $(\bv{\tilde A}_{i + b}^T \bv{\tilde A}_{i + b} + \lambda \bv{I})$. Using a trick introduced in \cite{spielman2011graph}, by applying a Johnson-Lindenstrauss random projection to the rows whose scores we are computing, we need just $\Oh(\log(1/\delta))$ system solves to compute constant factor approximations to the ridge scores with probability $1-\delta$. If we set $\delta = 1/\poly{n}$ then we can union bound over our whole stream, using this trick with each batch of $\Oh(d)$ input rows.
The batch probabilities will only be closer to the true ridge leverage scores than the non-batch probabilities and we will enjoy the same guarantees as $\textsc{Online-Sample}$.

Additionally, it turns out that with a simple trick, it is possible to reduce the memory usage of the algorithm by a factor of $\epsilon^{-2}$,
bringing it down to $\Oh(d \log d \log(\epsilon\norm{A}_2^2/\delta))$ (assuming the row sample is output to an output stream).
Note that this expression gets \emph{smaller} with $\epsilon$; hence we obtain a row sampling algorithm with memory complexity independent of desired multiplicative precision. The basic idea is that, instead of keeping all previously sampled rows in memory, we store a smaller set of rows that give a constant factor spectral approximation, still enough to give good estimates of the online ridge leverage scores.

This result is presented in the algorithm $\textsc{Slim-Sample}$ (Figure \ref{fig:slim-sample}) and Lemma \ref{lem:slim-sample}.
A particularly interesting consequence for graphs with polynomially bounded edge weights is:

\begin{corollary}
    \label{col:graph}
    Let $G$ be a simple graph on $d$ vertices, and $\epsilon \in (0,1)$.
    We can construct a $(1+\epsilon)$-sparsifier of $G$ of size $\Oh(d \log^2 d / \epsilon^2)$, using only $\Oh(d \log^2 d)$ working memory in the online model.
\end{corollary}
\Proof
This follows simply from applying Theorem \ref{thm:online-sample} with $\delta = \epsilon/\sigma_{min}^2(\bv A)$ and noting that the condition number of a graph on $d$ vertices whose edge weights are within a multiplicative $\poly{d}$ of each other is polynomial in $d$. So $\log(\epsilon \norm{\bv{A}}_2^2/\delta) = \log(\kappa^2(\bv A)) = \Oh(\log d)$.
\QED

We remark that the algorithm of Corollary \ref{col:graph} can be made to run in nearly linear time in the stream size. We combine $\textsc{Slim-Sample}$  with the batch processing idea described above. Because $\bv{A}$ is a graph, our matrix approximation is always a symmetric diagonally dominant matrix, with $\Oh(d)$ nonzero entries. We can solve systems in it in time $\tilde \Oh(d)$. Using the Johnson-Lindenstrauss random projection trick of \cite{spielman2011graph}, we can compute approximate ridge leverage scores for a batch of $\Oh(d)$ rows with failure probability polynomially small in $n$ in $\tilde \Oh(d \log n)$ time. Union bounding over the whole stream, we obtain nearly linear runtime.

To complement the row sampling results discussed above, 
we explore the limits of the proposed online setting.
In Section \ref{sec:bss} we present the algorithm $\textsc{Online-BSS}$, which obtains spectral approximations with $\Oh(d\log(\epsilon \norm{\bv{A}}_2^2/\delta)/\epsilon^2)$ rows in the online setting (with larger memory requirements than the simpler sampling algorithms).
Its analysis is given in Theorem \ref{thm:bss}.
In Section \ref{sec:lower}, we show that this number of samples is in fact the best achievable, up to constant factors (Theorem \ref{thm:lower_bound}). The $\log(\epsilon \norm{\bv{A}}_2^2/\delta)$ factor is truly the cost  of requiring rows to be selected in an online manner.

\section{Analysis of Sampling Schemes}
\label{sec:sampling}

We begin by bounding the sum of online $\lambda$-ridge leverage scores.
The intuition behind the proof of Theorem \ref{thm:online_scores} is that whenever we add a row with a large online leverage score to a matrix, we increase its determinant significantly, as follows from the matrix determinant lemma (Lemma \ref{lem:matdet}).
Thus we can reduce upper bounding the online leverage scores to bounding the matrix determinant.
\begin{lemma}[Matrix determinant lemma]
    \label{lem:matdet}
    Assume $\bv{S}$ is an invertible square matrix and $\bv{u}$ is a vector.
    Then
    \begin{align*}
        \det(\bv{S} + \bv{u}\bv{u}^T) = (\det \bv{S})(1 + \bv{u}^T\bv{S}^{-1}\bv{u}).
    \end{align*}
\end{lemma}

\Proofof{Theorem \ref{thm:online_scores}}
    By Lemma \ref{lem:matdet}, we have
    \begin{align*}
        \det (\bv{A}_{i+1}^T\bv{A}_{i+1} + \lambda \bv{I}) &= \det (\bv{A}_i^T \bv{A}_i + \lambda \bv{I}) \cdot \left(1 + \bv{a}_{i+1}^T (\bv{A}_i^T \bv{A}_i + \lambda \bv{I})^{-1} \bv{a}_{i+1}\right)\\
                                           &\geq \det (\bv{A}_i^T \bv{A}_i + \lambda \bv{I}) \cdot (1 + l_{i+1})\\
                                           &\geq \det (\bv{A}_i^T \bv{A}_i + \lambda \bv{I}) \cdot e^{l_{i+1} / 2}.
    \end{align*}
    Hence,
    \begin{align*}
        \det (\bv{A}^T \bv{A} + \lambda \bv{I}) &= \det (\bv{A}_n^T \bv{A}_n + \lambda \bv{I})\\
                               &\geq \det (\lambda \bv{I}) \cdot e^{\sum l_i / 2}\\
                               &= \lambda^d e^{\sum l_i / 2}.
    \end{align*}
    We have $\det (\bv{A}^T \bv{A} + \lambda \bv{I}) \leq (\|\bv{A}\|_2^2 + \lambda)^d$.
    Therefore
    \begin{align*}
        (\|\bv{A}\|_2^2+\lambda)^d &\geq \lambda^d e^{\sum l_i / 2}.
    \end{align*}
    Taking logarithms of both sides, we obtain
    \begin{align*}
          d \log (\|\bv{A}\|_2^2 + \lambda) &\geq d \log \lambda + \sum l_i / 2,\\
   \sum l_i &\leq 2 d \log (1 + \|\bv{A}\|_2^2/\lambda).
    \end{align*}
\QED

We now turn to analyzing the algorithm $\textsc{Online-Sample}$.
Because the samples taken by the algorithm are \emph{not} independent, we are not able to use a standard matrix Chernoff bound like the one in Theorem \ref{thm:leverage_sampling}.
However, we do know that whether we take row $i$ does not depend on later rows; thus, we are able to analyze the process as a martingale.
We will use a matrix version of the Freedman inequality given by Tropp.

\begin{theorem}[Matrix Freedman inequality \cite{tropp11}]
    \label{thm:matrix_freedman}
    Let $\bv{Y}_0, \bv{Y}_1, \ldots, \bv{Y}_n$ be a matrix martingale whose values are self-adjoint matrices with dimension $d$, and let $\bv{X}_1, \ldots, \bv{X}_n$ be the difference sequence.
    Assume that the difference sequence is uniformly bounded in the sense that
    \begin{align*}
        \norm{\bv{X}_k}_2 &\leq R\mbox{ almost surely, for } k = 1, \ldots, n.
    \end{align*}
    Define the predictable quadratic variation process of the martingale:
    \begin{align*}
        \bv{W}_k := \sum_{j=1}^k \expct{j - 1}{}{\bv{X}_j^2}\mbox{, for }k = 1, \ldots, n.
    \end{align*}
    Then, for all $\epsilon > 0$ and $\sigma^2 > 0$,
    \begin{align*}
        \prob{\norm{\bv{Y}_n}_2 \geq \epsilon\mbox{ and } \norm{\bv{W}_n}_2 \leq \sigma^2} &\leq d\cdot\exp\left(-\frac{-\epsilon^2/2}{\sigma^2+R\epsilon/3}\right)
    \end{align*}
\end{theorem}

We begin by showing that the output of $\textsc{Online-Sample}$ is in fact an approximation of $\bv{A}$, and that the approximate online leverage scores are lower bounded by the actual online leverage scores.

\begin{lemma}
    \label{lem:online-sample}
    After running $\textsc{Online-Sample}$, it holds with high probability that
    \begin{align*}
        (1-\epsilon)\bv{A}^T\bv{A} - \delta\bv{I} \preceq \bv{\tilde{A}}^T\bv{\tilde{A}} \preceq (1+\epsilon)\bv{A}^T\bv{A} + \delta\bv{I},
    \end{align*}
    and also
    \begin{align*}
        \tilde{l}_i \geq \bv{a}_i^T (\bv{A}^T\bv{A} + \lambda\bv{I})^{-1} \bv{a}_i
    \end{align*}
    for $i = 1, \ldots, n$.
\end{lemma}
\Proof
    Let
    \begin{align*}
        \bv{u}_i &:= (\bv{A}^T\bv{A} + \lambda \bv{I})^{-1/2} \bv{a}_i.
    \end{align*}
    We construct a matrix martingale $\bv{Y}_0, \bv{Y}_1, \ldots, \bv{Y}_n \in \mathbb{R}^{d\times d}$ with the difference sequence $\bv{X}_1, \ldots, \bv{X}_n$.
    Set $\bv {Y}_0 = \bv 0$. If $\norm{\bv{Y}_{i-1}}_2 \geq \epsilon$, we set $\bv{X}_i := \bv{0}$.
    Otherwise, let
    \begin{align*}
        \bv{X}_i &:=
        \begin{cases}
            (1/p_i - 1) \bv{u}_i \bv{u}_i^T &\mbox{ if $\bv{a}_i$ is sampled in $\bv{\tilde A}$,}\\
            - \bv{u}_i \bv{u}_i^T &\mbox{ otherwise.}
        \end{cases}
    \end{align*}
    Note that in this case we have
    \begin{align*}
        \bv{Y}_{i-1} = (\bv{A}^T\bv{A} + \lambda \bv{I})^{-1/2}(\bv{\tilde{A}}_{i-1}^T\bv{\tilde{A}}_{i-1} - \bv{A}_{i-1}^T\bv{A}_{i-1})(\bv{A}^T\bv{A} + \lambda \bv{I})^{-1/2}.
    \end{align*}
    Hence, since $\norm{\bv{Y}_{i-1}}_2 < \epsilon$, we have
    \begin{align*}
        \tilde{l}_i &= \min((1+\epsilon)\bv{a}_i^T (\bv{\tilde{A}}_{i-1}^T \bv{\tilde{A}}_{i-1} + \lambda \bv{I})^{-1} \bv{a}_i, 1) \\
                    &\geq \min((1+\epsilon)\bv{a}_i^T (\bv{A}_{i-1}^T \bv{A}_{i-1} + \lambda \bv{I} + \epsilon (\bv{A}^T\bv{A} + \lambda \bv{I}))^{-1} \bv{a}_i, 1) \\
                    &\geq \min((1+\epsilon)\bv{a}_i^T ((1+\epsilon)(\bv{A}^T \bv{A} + \lambda \bv{I}))^{-1} \bv{a}_i, 1) \\
                    &= \bv{a}_i^T (\bv{A}^T\bv{A} + \lambda\bv{I})^{-1} \bv{a}_i \\
                    &= \bv{u}_i^T\bv{u}_i,
    \end{align*}
    and so
    $
        p_i \geq \min(c \bv{u}_i^T \bv{u}_i, 1).
    $
    If $p_i = 1$, then $\bv{X}_i = 0$.
    Otherwise, we have $p_i \geq c \bv{u}_i^T \bv{u}_i$ and so
    \begin{align*}
        \|\bv{X}_i\|_2 &\leq 1 / c
    \end{align*}
    and
    \begin{align*}
        \expct{i-1}{}{\bv{X}_i^2} &\preceq p_i \cdot (1 / p_i - 1)^2 (\bv{u}_i \bv{u}_i^T)^2 + (1 - p_i) \cdot (\bv{u}_i \bv{u}_i^T)^2\\
                                &= (\bv{u}_i \bv{u}_i^T)^2 / p_i\\
                                &\preceq \bv{u}_i \bv{u}_i^T / c.
    \end{align*}
    And so, for the predictable quadratic variation process of the martingale $\{\bv{Y}_i\}$:
    \begin{align*}
        \bv{W}_i &:= \sum_{k=1}^i \expct{k-1}{}{\bv{X}_k^2},
    \end{align*}
    we have
    \begin{align*}
        \norm{\bv{W}_i}_2 \leq \left | \left |\sum_{k=1}^i \bv{u}_i \bv{u}_i^T / c \right | \right |_2 \le 1/c.
    \end{align*}

    Therefore by, Theorem \ref{thm:matrix_freedman}, we have
    \begin{align*}
        \prob{\norm{\bv{Y}_n}_2\geq \epsilon} &\leq d\cdot\exp\left(\frac{-\epsilon^2/2}{1/c + \epsilon/(3c)}\right)\\
                                              &\leq d\cdot\exp(-c\epsilon^2/4)\\
                                              &= 1/d.
    \end{align*}
    This implies that with high probability
    \begin{align*}
        \norm{(\bv{A}^T\bv{A} + \lambda \bv{I})^{-1/2}(\bv{\tilde{A}}^T\bv{\tilde{A}} + \lambda\bv{I})(\bv{A}^T\bv{A} + \lambda \bv{I})^{-1/2} - \bv{I}}_2 \leq \epsilon
    \end{align*}
    and so
    \begin{align*}
        (1-\epsilon)(\bv{A}^T\bv{A} + \lambda \bv{I}) &\preceq \bv{\tilde{A}}^T\bv{\tilde{A}} + \lambda\bv{I} \preceq (1+\epsilon)(\bv{A}^T\bv{A} + \lambda \bv{I}).
    \end{align*}
    Subtracting $\lambda\bv{I} = (\delta/\epsilon) \bv I$ from all sides, we get
    \begin{align*}
        (1-\epsilon)\bv{A}^T\bv{A} - \delta\bv{I} &\preceq \bv{\tilde{A}}^T\bv{\tilde{A}} \preceq (1+\epsilon)\bv{A}^T\bv{A} + \delta\bv{I}.
    \end{align*}
\QED

If we set $c$ in $\textsc{Online-Sample}$ to be proportional to $\log n$ rather than $\log d$, we would be able to take a union bound over all the rows and guarantee that with high probability all the approximate online leverage scores $\tilde{l}_i$ are close to true online leverage scores $l_i$.
Thus Theorem \ref{thm:online_scores} would imply that $\textsc{Online-Sample}$ only selects  $\Oh(d\log n \log(\|\bv{A}\|_2^2 / \lambda)/\epsilon^2)$ rows with high probability.

In order to remove the dependency on $n$, we have to sacrifice achieving close approximations to $l_i$ at every step.
Instead, we show that the \emph{sum} of the computed approximate online leverage scores is still small with high probability, using a custom Chernoff bound.

\begin{lemma}
    \label{lem:online-sample-fast}
    After running $\textsc{Online-Sample}$, it holds with high probability that
    \begin{align*}
        \sum_{i=1}^n \tilde{l}_i &= \Oh(d \log(\|\bv{A}\|_2^2 / \lambda)).
    \end{align*}
\end{lemma}
\Proof
    Define
    \begin{align*}
        \delta_i &:= \log \det(\bv{\tilde{A}}_i^T \bv{\tilde{A}}_i + \lambda \bv{I}) - \log \det(\bv{\tilde{A}}_{i-1}^T \bv{\tilde{A}}_{i-1} + \lambda \bv{I}).
    \end{align*}
    The proof closely follows the idea from the proof of Theorem \ref{thm:online_scores}.
    We will aim to show that large values of $\tilde{l}_i$ correlate with large values of $\delta_i$.
    However, the sum of $\delta_i$ can be bounded by the logarithm of the ratio of the determinants of $\bv{\tilde{A}}^T\bv{\tilde{A}} + \lambda\bv{I}$ and $\lambda\bv{I}$.
    First, we will show that $\expct{i-1}{}{\exp(\tilde{l}_i/8 - \delta_i)}$ is always at most $1$.
    We begin by an application of Lemma \ref{lem:matdet}.
    \begin{align*}
        \expct{i-1}{}{\exp(\tilde{l}_i/8 - \delta_i)} &= p_i \cdot e^{l_i/8} (1 + \bv{a}_i^T (\bv{\tilde{A}}_{i-1}^T \bv{\tilde{A}}_{i-1} + \lambda \bv{I})^{-1} \bv{a}_i / p_i)^{-1} + (1 - p_i) e^{l_i/8}\\
                                                    &\leq p_i \cdot (1 + l_i/4) (1 + \bv{a}_i^T (\bv{\tilde{A}}_{i-1}^T \bv{\tilde{A}}_{i-1} + \lambda \bv{I})^{-1} \bv{a}_i / p_i)^{-1} + (1 - p_i) (1 + l_i/4).
    \end{align*}
    If $c\tilde{l}_i < 1$, we have $p_i = c\tilde{l}_i$ and $\tilde{l}_i = (1+\epsilon)\bv{a}_i^T (\bv{\tilde{A}}_{i-1}^T \bv{\tilde{A}}_{i-1} + \lambda \bv{I})^{-1} \bv{a}_i$, and so:
    \begin{align*}
        \expct{i-1}{}{\exp(\tilde{l}_i/8 - \delta_i)} &\leq c\tilde{l}_i \cdot (1 + l_i/4) (1 + 1 / ((1+\epsilon)c))^{-1} + (1 - c\tilde{l}_i) (1 + l_i/4)\\
                                                    &= (1 + l_i/4) (cl_i (1 + 1/((1+\epsilon)c))^{-1} + 1 - cl_i)\\
                                                    &\leq (1 + l_i/4) (1 + cl_i (1 - 1/(4c) - 1))\\
                                                    &= (1 + l_i/4) (1 - l_i/4)\\
                                                    &\leq 1.
    \end{align*}
    Otherwise, we have $p_i = 1$ and so:
    \begin{align*}
        \expct{i-1}{}{\exp(\tilde{l}_i/8 - \delta_i)} &\leq (1 + l_i/4) (1 + \bv{a}_i^T (\bv{\tilde{A}}_{i-1}^T \bv{\tilde{A}}_{i-1} + \lambda \bv{I})^{-1} \bv{a}_i)^{-1} \\
                                                    &\leq (1 + l_i/4) (1 + l_i)^{-1} \\
                                                    &\leq 1.
    \end{align*}
    We will now analyze the expected product of $\exp(\tilde{l}_i/8 - \delta_i)$ over the first $k$ steps.
    We group the expectation over the first $k$ steps into one over the first $k-1$ steps, aggregating the expectation for the last step by using one-way independence.
    For $k \geq 1$ we have
    \begin{align*}
        \expct{}{}{\exp\left(\sum_{i=1}^k \tilde{l}_i/8 - \delta_i\right)} &= \expct{}{\mbox{first $k-1$ steps}}{\exp\left(\sum_{i=1}^{k-1} \tilde{l}_i/8 - \delta_i\right)\expct{k-1}{}{\exp(\tilde{l}_k/8 - \delta_k)}}\\
                                                                         &\leq \expct{}{}{\exp\left(\sum_{i=1}^{k-1} \tilde{l}_i/8 - \delta_i\right)},
    \end{align*}
    and so by induction on $k$
    \begin{align*}
        \expct{}{}{\exp\left(\sum_{i=1}^n \tilde{l}_i/8 - \delta_i\right)} &\leq 1.
    \end{align*}
    Hence by Markov's inequality
    \begin{align*}
        \prob{\sum_{i=1}^n \tilde{l}_i > 8d + 8\sum_{i=1}^n \delta_i} &\leq e^{-d}.
    \end{align*}
    By Lemma \ref{lem:online-sample}, with high probability we have
    \begin{align*}
        \bv{\tilde{A}}^T\tilde{\bv{A}} + \lambda \bv{I} &\preceq (1+\epsilon)(\bv{A}^T\bv{A} + \lambda \bv{I}).
    \end{align*}
    We also have with high probability
    \begin{align*}
        \det (\bv{\tilde{A}}^T\tilde{\bv{A}} + \lambda \bv{I}) &\leq (1+\epsilon)^d(\norm{A}_2^2 + \lambda)^d,\\
        \log \det (\bv{\tilde{A}}^T\tilde{\bv{A}} + \lambda \bv{I}) &\leq d(1 + \log(\norm{A}_2^2 + \lambda)).
    \end{align*}
    Hence, with high probability it holds that
    \begin{align*}
        \sum_{i=1}^n \delta_i &= \log \det (\bv{\tilde{A}}^T\tilde{\bv{A}} + \lambda \bv{I}) - d \log(\lambda)\\
                              &\leq d(1 + \log(\norm{A}_2^2 + \lambda) - \log(\lambda))\\
                              &= d(1 + \log(1 + \norm{A}_2^2 / \lambda)).
    \end{align*}
    And so, with high probability,
    \begin{align*}
        \sum_{i=1}^n \tilde{l}_i &\leq 8d + 8\sum_{i=1}^n \delta_i\\
                                 &\leq 9d + 8d\log(1 + \norm{A}_2^2 / \lambda)\\
                                 &= \Oh(d \log(\norm{A}_2^2 / \lambda)).
    \end{align*}
\QED

\Proofof{Theorem \ref{thm:online-sample}}
The thesis follows immediately from Lemmas \ref{lem:online-sample} and \ref{lem:online-sample-fast}.
\QED

We now consider a simple modification of $\textsc{Online-Sample}$ that removes dependency on $\epsilon$ from the working memory usage with no additional cost.

\begin{figure}[ht]
\noindent
\centering
\fbox{
\begin{minipage}{6in}
    \noindent $\bv{\tilde{A}} = \textsc{Slim-Sample} (\bv{A}, \epsilon, \delta)$,
    where $\bv{A}$ is an $n \times d$ matrix with rows $\bv{a}_1, \ldots, \bv{a}_n$, $\epsilon \in (0, 1)$, $\delta > 0$.
\begin{enumerate}
\item Set $\lambda := \delta / \epsilon$, $c := 8\log d / \epsilon^2$.
\item Let $\bv{\tilde{A}}_0$ be a $0 \times d$ matrix.
\item Let $\tilde{l}_1, \ldots, \tilde{l}_n$ be the approximate online leverage scores computed by an independent instance of $\textsc{Online-Sample}(\bv{A}, 1/2, \delta/(2\epsilon))$.
\item For $i = 1, \ldots, n$:
    \begin{enumerate}
        \item Let $p_i := \min(c\tilde{l}_i, 1)$.
        \item Set
            $
            \bv{\tilde{A}}_i := 
            \begin{cases}
            \begin{bmatrix} \bv{\tilde{A}}_{i-1}\\\bv{a}_i / \sqrt{p_i}\end{bmatrix} &\mbox{ with probability $p_i$,} \vspace{0.2cm}\\
                \bv{\tilde{A}}_{i-1} &\mbox{ otherwise.}\\
            \end{cases}
            $
    \end{enumerate}
\item Return $\bv{\tilde{A}} := \bv{\tilde{A}}_n$.
\end{enumerate}
\end{minipage}
}
\caption{The low-memory online sampling algorithm}
\label{fig:slim-sample}
\end{figure}

\begin{lemma}
    \label{lem:slim-sample}
    Let $\bv{\tilde{A}}$ be the matrix returned by $\textsc{Slim-Sample}(\bv{A}, \epsilon, \delta)$.
    Then, with high probability,
    \begin{align*}
        (1-\epsilon)\bv{A}^T\bv{A} - \delta\bv{I} \preceq \bv{\tilde{A}}^T\bv{\tilde{A}} \preceq (1+\epsilon)\bv{A}^T\bv{A} + \delta\bv{I},
    \end{align*}
    and the number of rows in $\bv{\tilde{A}}$ is $\Oh(d \log d \log(\epsilon\norm{\bv{A}}_2^2/\delta)/\epsilon^2)$.

    Moreover, with high probability the algorithm $\textsc{Slim-Sample}$'s memory requirement is dominated by storing $\Oh(d \log d \log(\epsilon\norm{\bv{A}}_2^2/\delta))$ rows of $\bv{A}$.
\end{lemma}

\Proof
As the samples are independent, the thesis follows from Theorem \ref{thm:leverage_sampling} and Lemmas \ref{lem:online-sample} and \ref{lem:online-sample-fast}.
\QED

\section{Asymptotically Optimal Algorithm}
\label{sec:bss}
In addition to sampling by online leverage scores, there is also a variant of the ``BSS'' method \cite{batson2012twice} that applies in our setting.  Like the original \cite{batson2012twice}, this approach removes the $\log d$ factor from the row count of the output spectral approximation, matching the lower bound for online sampling given in Theorem \ref{thm:lower_bound}.  

Unlike \cite{batson2012twice} itself, our algorithm is randomized -- it is similar to, and inspired by, the randomized version of BSS from \cite{lee2015linear}, especially the simpler ``Algorithm 1'' from that paper (the main difference from that is considering each row separately).  In fact, this algorithm is of the same form as the basic sampling algorithm, in that when each row comes in, a probability $p_i$ is assigned to it, and it is kept (and rescaled) with probability $p_i$ and rejected otherwise.  The key difference is the definition of the $p_i$.

There are also some differences in the nature of the algorithm and its guarantees.  Notably, the $p_i$ cannot be computed solely based on the row sample output so far--it is necessary to ``remember'' the entire matrix given so far.  This means that the BSS method is not memory efficient, using $O(d^2)$ space. Additionally, online leverage score sampling gives bounds on both the size of the output spectral approximation and its accuracy with high probability.  In contrast, this method gives an \emph{expected} bound on the output size, while it \emph{never} fails to output a correct spectral approximation.  Note that these guarantees are essentially the same as those in the appendix of \cite{lee2015linear}.

One may, however, improve the memory dependence in some cases simply by running it on the output stream of the online leverage score sampling method.  This reduces the storage cost to the size of that spectral approximation.  The BSS method still does not produce an actual space \emph{savings} (in particular, there is a still a $\log d$ factor in space), but it does reduce the number of rows in the output stream while only blowing up the space usage by $O(1/\epsilon^2)$ (due to requiring the storage of an $\epsilon$-quality approximation rather than only $O(1)$).

The BSS method maintains two matrices, $\bv{B}^U_i$ and $\bv{B}^L_i$, acting as upper and lower ``barriers''.  The current spectral approximation will always fall between them:
\begin{equation*}
\bv{B}^L_i \prec \bv{\tilde{A}}_i^T \bv{\tilde{A}}_i^T \prec \bv{B}^U_i.
\end{equation*}
This guarantee, at the end of the algorithm, will ensure that $\bv{\tilde{A}}$ is a valid approximation.

Below, we give the actual BSS algorithm and its performance guarantees.
\begin{theorem}
\leavevmode
\label{thm:bss}
\begin{enumerate}
\item
The online BSS algorithm always outputs $\tilde{A}$ such that
\begin{equation*}
(1-\epsilon) \bv{A}^T \bv{A} - \delta \bv{I} \prec \bv{\tilde{A}}^T \bv{\tilde{A}}^T \prec (1+\epsilon) \bv{A}^T \bv{A} + \delta \bv{I}
\end{equation*}

\item
The probability that a row $\bv{a}_i$ is included in $\bv{\tilde{A}}$ is at most $\frac{8}{\epsilon^2} l_i$, where $l_i$ is the online $\frac{2 \delta}{\epsilon}$-ridge leverage score of $\bv{a}_i$. That is $l_i = \min(\bv{a}_i^T \left ( \bv{A}_i^T \bv{A}_i + \frac{2 \delta}{\epsilon} I \right )^{-1} \bv{a}_i, 1)$.  The expected number of rows in $\bv{\tilde{A}}$ is thus at most $\frac{8}{\epsilon^2} \sum_{i=1}^n l_i =\Oh(d \log (\epsilon\norm{\bv A}_2^2/\delta) / \epsilon^2)$.
\end{enumerate}
\end{theorem}
\begin{figure}[ht]
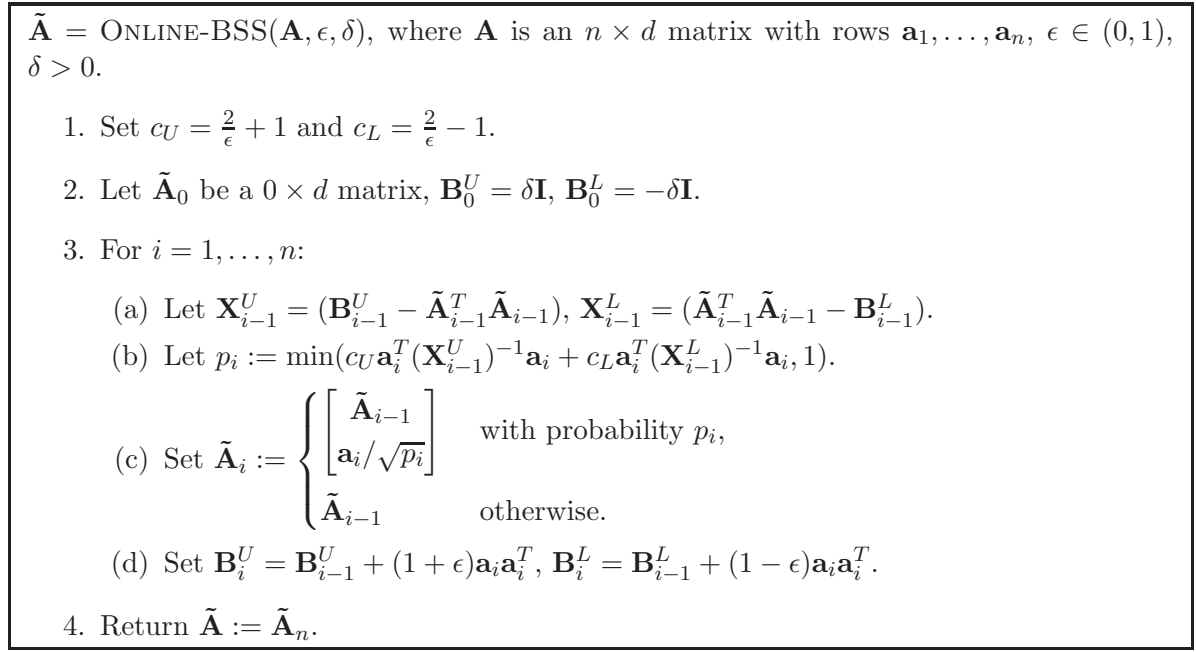

\noindent
\centering
\fbox{
\begin{minipage}{6in}
    \noindent $\bv{\tilde{A}} = \textsc{Online-BSS} (\bv{A}, \epsilon, \delta)$,
    where $\bv{A}$ is an $n \times d$ matrix with rows $\bv{a}_1, \ldots, \bv{a}_n$, $\epsilon \in (0, 1)$, $\delta > 0$.
\begin{enumerate}
\item Set $c_U = \frac{2}{\epsilon}+1$ and $c_L = \frac{2}{\epsilon}-1$.
\item Let $\bv{\tilde{A}}_0$ be a $0 \times d$ matrix, $\bv{B}^U_0 = \delta \bv{I}$, $\bv{B}^L_0 = -\delta \bv{I}$.
\item For $i = 1, \ldots, n$:
    \begin{enumerate}
    	\item Let $\bv{X}^U_{i-1} = (\bv{B}^U_{i-1}-\bv{\tilde{A}}_{i-1}^T \bv{\tilde{A}}_{i-1})$, $\bv{X}^L_{i-1} = (\bv{\tilde{A}}_{i-1}^T \bv{\tilde{A}}_{i-1} - \bv{B}^L_{i-1})$.
        \item Let $p_i := \min(c_U \bv{a}_i^T (\bv{X}^U_{i-1})^{-1} \bv{a}_i + c_L \bv{a}_i^T (\bv{X}^L_{i-1})^{-1} \bv{a}_i, 1)$.
        \item Set
            $
            \bv{\tilde{A}}_i := 
            \begin{cases}
            \begin{bmatrix} \bv{\tilde{A}}_{i-1}\\\bv{a}_i / \sqrt{p_i}\end{bmatrix} &\mbox{ with probability $p_i$,} \vspace{0.2cm}\\
                \bv{\tilde{A}}_{i-1} &\mbox{ otherwise.}\\
            \end{cases}
            $
        \item Set $\bv{B}^U_i = \bv{B}^U_{i-1} + (1+\epsilon) \bv{a}_i \bv{a}_i^T$, $\bv{B}^L_i = \bv{B}^L_{i-1} + (1-\epsilon) \bv{a}_i \bv{a}_i^T$.
    \end{enumerate}
\item Return $\bv{\tilde{A}} := \bv{\tilde{A}}_n$.
\end{enumerate}
\end{minipage}
}
\caption{The Online BSS Algorithm}
\label{fig:online-bss}
\end{figure}

\Proofof{Theorem \ref{thm:bss} part 1}
We first note the basic invariant that $\bv{X}^U_i$ and $\bv{X}^L_i$ always remain positive definite--or equivalently,
\begin{equation*}
\bv{B}^L_i \prec \bv{\tilde{A}}_i^T \bv{\tilde{A}}_i^T \prec \bv{B}^U_i.
\end{equation*}

We may prove this by induction on $i$.  The base case follows from the initialization of $\bv{\tilde{A}}_0$, $\bv{B}^U_0$ and $\bv{B}^L_0$.  For each successive step, we consider two possibilities.

The first is that $p_i = 1$.  In that case, $\bv{\tilde{A}}^T \bv{\tilde{A}}$ always increases by exactly $\bv{a}_i \bv{a}_i^T$, $\bv{B}^U$ by $(1+\epsilon) \bv{a}_i \bv{a}_i^T$ and $\bv{B}^L$ by $(1-\epsilon) \bv{a}_i \bv{a}_i^T$.  Thus $\bv{X}^U$ and $\bv{X}^L$ increase by exactly $\epsilon \bv{a}_i \bv{a}_i^T$, which is positive semidefinite, and so remain positive definite.

In the other case, $p_i < 1$.  Now, $\bv{X}^U$ decreases by at most the increase in $\bv{\tilde{A}}_i^T \bv{\tilde{A}}_i^T$, or
\begin{equation*}
\bv{M}_i = \frac{\bv{a}_i \bv{a}_i^T}{p}.
\end{equation*}
Since $c_U > 1$, $p > \bv{a}_i^T (\bv{X}^U_{i-1})^{-1} \bv{a}_i$, so $\bv{a}_i \bv{a}_i^T \prec p \bv{X}^U_{i-1}$ and $\bv{M}_i \prec \bv{X}^U_{i-1}$.  Subtracting this then must leave $\bv{X}^U$ positive definite.
Similarly, $\bv{X}^L$ decreases by at most the increase in $\bv{B}^L$, which is $(1-\epsilon) \bv{a}_i \bv{a}_i^T \prec \bv{a}_i \bv{a}_i^T$.  Since $c_L > 1$ and $p < 1$, $\bv{a}_i^T (\bv{X}^L_{i-1})^{-1} \bv{a}_i < 1$, and $\bv{a}_i \bv{a}_i^T \prec \bv{X}^L_{i-1}$.  Subtracting this similarly leaves $\bv{X}^L$ positive definite.
Finally, we note that
\begin{align*}
\bv{B}^U_n &= (1+\epsilon) \bv{A}^T \bv{A} + \delta \bv{I} \\
\bv{B}^L_n &= (1-\epsilon) \bv{A}^T \bv{A} - \delta \bv{I}.
\end{align*}
This gives the desired result.
\QED

To prove part 2, we will use quantities of the form $\bv{v}^T \bv{X}^{-1} \bv{v}$.  We will need a lemma describing how this behaves under a random rank-1 update:
\begin{lemma}
\label{lem:rankoneupdate}
Given a positive definite matrix $\bv{X}$, two vectors $\bv{u}$ and $\bv{v}$, two multipliers $a$ and $b$ and a probability $p$, define the random variable $\bv{X}'$ to be $X - a \bv{u} \bv{u}^T$ with probability $p$ and $X - b \bv{u} \bv{u}^T$ otherwise.  Then if $\bv{u}^T \bv{X}^{-1} \bv{u} = 1$,
\begin{equation*}
\expct{}{}{\bv{v}^T \bv{X}'^{-1} \bv{v} - \bv{v}^T \bv{X}^{-1} \bv{v}] = (\bv{v^T} \bv{X}^{-1} \bv{u})^2 \frac{pa + (1-p)b - ab}{(1-a)(1-b)}}.
\end{equation*}
\end{lemma}
\Proof
We apply the Sherman-Morrison formula to each of the two possibilities (subtracting $a \bv{u} \bv{u}^T$ and $b \bv{u} \bv{u}^T$ respectively).  These give $\bv{X}'$ values of respectively
\begin{equation*}
\bv{X}^{-1} + a \frac{\bv{X}^{-1} \bv{u} \bv{u}^T \bv{X}^{-1}}{1 - a \bv{u}^T \bv{X}^{-1} \bv{u}} = \bv{X}^{-1} + \frac{a}{1-a} \bv{X}^{-1} \bv{u} \bv{u}^T \bv{X}^{-1}
\end{equation*}
and
\begin{equation*}
\bv{X}^{-1} + b \frac{\bv{X}^{-1} \bv{u} \bv{u}^T \bv{X}^{-1}}{1 - b \bv{u}^T \bv{X}^{-1} \bv{u}} = \bv{X}^{-1} + \frac{b}{1-b} \bv{X}^{-1} \bv{u} \bv{u}^T \bv{X}^{-1}.
\end{equation*}
The values of $\bv{v}^T \bv{X}'^{-1} \bv{v} - \bv{v}^T \bv{X}^{-1} \bv{v}$ are then respectively
\begin{equation*}
\frac{a}{1-a} \bv{v}^T \bv{X}^{-1} \bv{u} \bv{u}^T \bv{X}^{-1} \bv{v} = (\bv{v^T} \bv{X}^{-1} \bv{u})^2 \frac{a}{1-a}
\end{equation*}
and
\begin{equation*}
\frac{b}{1-b} \bv{v}^T \bv{X}^{-1} \bv{u} \bv{u}^T \bv{X}^{-1} \bv{v} = (\bv{v^T} \bv{X}^{-1} \bv{u})^2 \frac{b}{1-b}.
\end{equation*}
Combining these gives the stated result.
\QED

\Proofof{Theorem \ref{thm:bss} part 2}
First, we introduce some new matrices to help in the analysis:
\begin{align*}
\bv{C}^U_{i,j} &= \delta \bv{I} + \frac{\epsilon}{2} \bv{A}_i^T \bv{A}_i + \left ( 1 + \frac{\epsilon}{2} \right ) \bv{A}_j^T \bv{A}_j \\
\bv{C}^L_{i,j} &= -\delta \bv{I} - \frac{\epsilon}{2} \bv{A}_i^T \bv{A}_i + \left ( 1 - \frac{\epsilon}{2} \right ) \bv{A}_j^T \bv{A}_j.
\end{align*}
Note that $\bv{C}^U_{i,i} = \bv{B}^U_i$, $\bv{C}^L_{i,i} = \bv{B}^L_i$, and for $j \leq i$, $\bv{C}^U_{i,j} \succeq \bv{B}^U_j$ and $\bv{C}^L_{i,j} \preceq \bv{B}^L_j$.
We can then define:
\begin{align*}
\bv{Y}^U_{i,j} &= \bv{C}^U_{i,j} - \bv{\tilde{A}}_j^T \bv{\tilde{A}}_j \\
\bv{Y}^L_{i,j} &= \bv{\tilde{A}}_j^T \bv{\tilde{A}}_j - \bv{C}^L_{i,j}.
\end{align*}
We then have, similarly, $\bv{Y}^U_{i,i} = \bv{X}^U_i$, $\bv{Y}^L_{i,i} = \bv{X}^L_i$, and for $j \leq i$, $\bv{Y}^U_{i,j} \succeq \bv{X}^U_j$ and $\bv{Y}^L_{i,j} \succeq \bv{X}^L_j$.

We will assume that $l_i < 1$, since otherwise the claim is immediate (as probabilities cannot exceed 1).  Now, note that
\begin{align*}
\bv{a}_i^T (\bv{Y}^U_{i,0})^{-1} \bv{a}_i &= \bv{a}_i^T (\bv{Y}^L_{i,0})^{-1} \bv{a}_i \\
&= \bv{a}_i^T \left ( \frac{\epsilon}{2} \bv{A}_i^T \bv{A}_i + \delta I \right )^{-1} \bv{a}_i \\
&= \frac{2}{\epsilon} \left ( \bv{A}_i^T \bv{A}_i + \frac{2 \delta}{\epsilon} I \right )^{-1} \bv{a}_i \\
&= \frac{2}{\epsilon} l_i.
\end{align*}

Next, we will aim to show that for $j < i-1$,
\begin{align*}
\expct{}{}{\bv{a}_i^T \bv{Y}^U_{i-1,j+1} \bv{a}_i} &\leq \expct{}{}{\bv{a}_i^T \bv{Y}^U_{i-1,j} \bv{a}_i} \\
\expct{}{}{\bv{a}_i^T \bv{Y}^L_{i-1,j+1} \bv{a}_i} &\leq \expct{}{}{\bv{a}_i^T \bv{Y}^L_{i-1,j} \bv{a}_i}
\end{align*}

In particular, we will simply show that conditioned on any choices for the first $j$ rows, the expected value of $\bv{a}_i^T \bv{Y}^U_{i-1,j+1} \bv{a}_i$ is no larger than $\bv{a}_i^T \bv{Y}^U_{i-1,j} \bv{a}_i$, and analogously for $\bv{Y}^L$.

Similar to the proof of part 1, we separately consider the case where $p_{j+1} = 1$.  In that case, the positive semidefinite matrix $\frac{\epsilon}{2} \bv{a}_j \bv{a}_j^T$ is simply added to $\bv{Y}^U$ and $\bv{Y}^L$.  Adding this can only decrease the values of $\bv{a}_i^T \bv{Y}^U \bv{a}_i$ and $\bv{a}_i^T \bv{Y}^L \bv{a}_i$.

The $p_{j+1} < 1$ case is more tricky.  Here, we define the vector $\bv{w}_{j+1} = \frac{\bv{a}_{j+1}}{\sqrt{p_{j+1}}}$.  Importantly
\begin{gather*}
p_{j+1} \geq c_U \bv{a}_{j+1}^T (\bv{X}^U_j)^{-1} \bv{a}_{j+1} \geq c_U \bv{a}_{j+1}^T (\bv{Y}^U_{i-1,j})^{-1} \bv{a}_{j+1} \\
p_{j+1} \geq c_L \bv{a}_{j+1}^T (\bv{X}^L_j)^{-1} \bv{a}_{j+1} \geq c_L \bv{a}_{j+1}^T (\bv{Y}^L_{i-1,j})^{-1} \bv{a}_{j+1}.
\end{gather*}

This means that
\begin{align*}
\bv{w}_{j+1}^T (\bv{Y}^U_{i-1,j})^{-1} \bv{w}_{j+1}^T &\leq \frac{1}{c_U} \\
\bv{w}_{j+1}^T (\bv{Y}^L_{i-1,j})^{-1} \bv{w}_{j+1}^T &\leq \frac{1}{c_L}.
\end{align*}

Now, we additionally define
\begin{align*}
s^U_{j+1} &= \bv{w}_{j+1}^T (\bv{Y}^U_{i-1,j})^{-1} \bv{w}_{j+1}^T \\
s^L_{j+1} &= \bv{w}_{j+1}^T (\bv{Y}^L_{i-1,j})^{-1} \bv{w}_{j+1}^T \\
\bv{u}^U_{j+1} &= \frac{\bv{w}_{j+1}}{\sqrt{s^U_{j+1}}} \\
\bv{u}^L_{j+1} &= \frac{\bv{w}_{j+1}}{\sqrt{s^L_{j+1}}}.
\end{align*}

We then deploy Lemma \ref{lem:rankoneupdate} to compute the expectations.  For the contribution from the upper barrier, we use $\bv{X} = \bv{Y}^U_{i-1,j}$, $\bv{u} = \bv{u}^U_{j+1}$, $\bv{v} = \bv{a}_i^T$, $a = -s^U_{j+1} (1 - p_{j+1} (1 + \epsilon / 2))$, $b = s^U_{j+1} p_{j+1} (1 + \epsilon / 2)$, $p = p_{j+1}$.  For the lower barrier, we use $\bv{X} = \bv{Y}^L_{i-1,j}$, $\bv{u} = \bv{u}^L_{j+1}$, $\bv{v} = \bv{a}_i^T$, $a = s^L_{j+1} (1 - p_{j+1} (1 - \epsilon / 2))$, $b = -s^L_{j+1} p_{j+1} (1 - \epsilon / 2)$, $p = p_{j+1}$.  In both cases we can see that the numerator of the expected change is nonpositive.
Finally, this implies that the probability that row $i$ is sampled is
\begin{align*}
\expct{}{}{p_i} &= c_U \expct{}{}{\bv{a}_i^T (\bv{X}^U_{i-1})^{-1} \bv{a}_i} + c_L \expct{}{}{\bv{a}_i^T (\bv{X}^L_{i-1})^{-1} \bv{a}_i} \\
&= c_U \expct{}{}{\bv{a}_i^T (\bv{Y}^U_{i-1,i-1})^{-1} \bv{a}_i} + c_L \expct{}{}{\bv{a}_i^T (\bv{Y}^L_{i-1,i-1})^{-1} \bv{a}_i} \\
&\leq c_U \expct{}{}{\bv{a}_i^T (\bv{Y}^U_{i-1,0})^{-1} \bv{a}_i} + c_L \expct{}{}{\bv{a}_i^T (\bv{Y}^L_{i-1,0})^{-1} \bv{a}_i} \\
&= \frac{2}{\epsilon} (c_U + c_L) l_i \\
&= \frac{8}{\epsilon^2} l_i
\end{align*}
as desired.
\QED

\section{Matching Lower Bound}
\label{sec:lower}

Here we show that the row count obtained by Theorem \ref{thm:bss} is in fact optimal. While it is possible to obtain a spectral approximation with $\Oh(d/\epsilon^2)$ rows in the offline setting, online sampling always incurs a loss of $\Omega \left (\log(\epsilon\norm{\bv A}^2_2/\delta) \right)$ and must sample $\Omega \left (\frac{d\log(\epsilon\norm{\bv A}^2_2/\delta)}{\epsilon^2} \right)$ rows.

\begin{theorem}
	\label{thm:lower_bound}
	Assume that $\epsilon \norm{\bv{A}}_2^2 \ge c_1\delta$ and $\epsilon \ge c_2/\sqrt{d}$, for fixed constants $c_1$ and $c_2$.
	Then any algorithm that selects rows in an online manner and outputs a spectral approximation to $\bv{A}^T\bv{A}$ with $(1+\epsilon)$ multiplicative error and $\delta$ additive error with probability at least $1/2$ must sample $\Omega \left (\frac{d\log(\epsilon\norm{\bv A}^2_2/\delta)}{\epsilon^2} \right)$ rows of $\bv{A}$ in expectation.
\end{theorem}

Note that the lower bounds we assume on $\epsilon \norm{\bv{A}}_2^2$ and $\epsilon$ are very minor. They just ensure that $\log(\epsilon\norm{\bv A}^2_2/\delta) \ge 1$ and that $\epsilon$ is not so small that we can essentially sample all rows of $\bv{A}$.

\Proof
We apply Yao's minimax principle, constructing, for any large enough $M$, a distribution on inputs $\bv{A}$ with $\norm{\bv{A}}_2^2 \le M$ for which any deterministic online row selection algorithm that succeeds with probability at least $1/2$ must output  $\Omega \left (\frac{d\log(\epsilon M/\delta)}{\epsilon^2} \right)$ rows in expectation. The best randomized algorithm that works with probability $1/2$ on any input matrix with $\norm{\bv{A}}_2^2 \le M$ therefore must select at least $\Omega \left (\frac{d\log(\epsilon M/\delta)}{\epsilon^2} \right)$ rows in expectation on the worst case input, giving us the theorem.

Our distribution is as follows. We select an integer $N$ uniformly at random from $[1, \log(M\epsilon/\delta)]$. We then stream in the vertex edge incidence matrices of $N$ complete graphs on $d$ vertices. We double the weight of each successive graph. Intuitively, spectrally approximating a complete graph requires selecting $\Omega(d/\epsilon^2)$ edges \cite{batson2012twice} (as long as $\epsilon \ge c_2/\sqrt{d}$ for some fixed constant $c_2$). Each time we stream in a new graph with double the weight, we force the algorithm to add $\Omega(d/\epsilon^2)$  more edges to its output, eventually forcing it to output $\Omega(d/\epsilon^2 \cdot N)$ edges -- $\Omega(d\log(M\epsilon/\delta)/\epsilon^2)$ in expectation.

Specifically, let $\bv{K}_d$ be the ${d \choose 2} \times d$ vertex edge incidence matrix of the complete graph on $d$ vertices. $\bv{K}_d^T\bv{K}_d$ is the Laplacian matrix of the complete graph on $d$ vertices. We weight the first graph so that its Laplacian has all its nonzero eigenvalues equal to $\delta/\epsilon$. (That is, each edge has weight $\frac{\delta}{d\epsilon}$). In this way, even if we select $N = \lfloor \log(M\epsilon/\delta) \rfloor$ we will have overall $\norm{\bv{A}}^2_2 \le \delta/\epsilon + 2\delta/\epsilon + ... 2^{\lfloor \log(M\epsilon/\delta) \rfloor-1} \delta/\epsilon \le M$.

Even if $N = 1$, all nonzero eigenvalues of $\bv{A}^T\bv{A}$ are at least $\delta/\epsilon$, so achieving $(1+\epsilon)$ multiplicative error and $\delta \bv{I}$ additive error is equivalent to achieving $(1+2\epsilon)$ multiplicative error. $\bv{A}^T\bv{A}$ is a graph Laplacian so has a null space. However, as all rows are orthogonal to the null space, achieving additive error $\delta \bv{I}$ is equivalent to achieving additive error $\delta \bv{I}_r$ where $\bv{I}_r$ is the identity projected to the span of $\bv{A}^T\bv{A}$. $\delta \bv{I}_r \preceq \epsilon \bv{A}^T\bv{A}$ which is why we must achieve $(1+2\epsilon)$ multiplicative error.

In order for a deterministic algorithm to be correct with probability $1/2$ on our distribution, it must be correct for at least $1/2$ of our $\lfloor \log(M\epsilon/\delta) \rfloor$ possible choices of $N$.

Let $i$ be the lowest choice of $N$ for which the algorithm is correct. By the lower bound of \cite{batson2012twice}, the algorithm must output  $\Omega(d/\epsilon^2)$ rows of $\bv{A}_i$ to achieve a $(1+2\epsilon)$ multiplicative factor spectral approximation. Here $\bv{A}_i$ is the input consisting of the vertex edge incidence matrices of $i$ increasingly weighted complete graphs. Call the output on this input $\bv{\tilde A}_i$.
Now let $j$ be the second lowest choice of $N$ on which the algorithm is correct. Since the algorithm was correct on $\bv{A}_i$ to within a multiplicative $(1+2\epsilon)$, to be correct on $\bv{A}_j$, 
it must output a set of edges $\bv{\tilde A}_j$ such that 
\begin{align*}
(\bv{A}_j^T \bv{A}_j- \bv{A}_i^T \bv{A}_i) - 4 \epsilon \bv{A}_j^T \bv{A}_j \preceq \bv{\tilde A}_j^T\bv{\tilde A}_j - \bv{\tilde A}_i^T\bv{\tilde A}_i \preceq (\bv{A}_j^T \bv{A}_j- \bv{A}_i^T \bv{A}_i) + 4 \epsilon \bv{A}_j^T \bv{A}_j.
\end{align*}

Since we double each successive copy of the complete graph, $\bv{A}_j^T \bv{A}_j \preceq 2(\bv{A}_j^T \bv{A}_j- \bv{A}_i^T \bv{A}_i)$. So, $\bv{\tilde A}_j^T\bv{\tilde A}_j - \bv{\tilde A}_i^T\bv{\tilde A}_i$ must be a $1+8\epsilon$ spectral approximation to the true difference $\bv{A}_j^T \bv{A}_j - \bv{A}_i^T \bv{A}_i$. Noting that this difference is itself just a weighting of the complete graph, by the lower bound in \cite{batson2012twice} the algorithm must select $\Omega(d/\epsilon^2)$ additional edges between the $i^{th}$ and $j^{th}$ input graphs. Iterating this argument over all $\lfloor \log(M\epsilon/\delta) \rfloor/2$ inputs on which the algorithm must be correct, it must select a total of $\Omega(d\log(M\epsilon/\delta)/\epsilon^2)$ edges in expectation over all inputs. 
\QED

\section{Future Work}

An obvious open question arising from our work is if one can prove that the algorithm of \cite{kelner2013spectral} works despite dependencies arising due to the row pruning step. By operating in the online setting, our algorithm avoids row pruning, and hence is able to skirt these dependencies, as the probability that a row is sampled only depends on earlier rows in the stream. However, because the streaming setting offers the potential for sampling fewer rows than in the online case, obtaining a rigorous proof of \cite{kelner2013spectral} would be very interesting.

While our work focuses on spectral approximation, variants on (ridge) leverage score sampling and the BSS algorithm are also used to solve low-rank approximation problems, including column subset selection \cite{boutsidis2014optimal,cohen2015ridge} and projection-cost-preserving sketching \cite{cohen2014dimensionality,cohen2015ridge}. Compared with spectral approximation, there is less work on streaming sampling for low-rank approximation, and understanding how online algorithms may be used in this setting would an interesting extension of our work.

\section{Acknowledgments}
The authors would like to thank Kenneth Clarkson, Jonathan Kelner, Gary Miller, Christopher Musco and Richard Peng for helpful discussions and comments.

Cameron Musco and Jakub Pachocki both acknowledge the Gene Golub SIAM Summer School program on Randomization in Numerical Linear Algebra, where work on this project was initiated.

\bibliographystyle{alpha}
\bibliography{online}

\appendix

\end{document}